\documentclass[nofootinbib]{revtex4}
\usepackage[utf8]{inputenc}
\usepackage{amsmath,amssymb,graphicx}

\newcommand{\angstr}{\rm\AA}

\begin{document}

\title{Heisenberg models with minimal number of parameters\\for two-dimensional magnetic crystals}

\author{K.~Zberecki$^1$, M.~Wilczyński$^1$, and M.~Wierzbicki$^1$} 
\affiliation{$^1$Faculty of Physics, Warsaw University of Technology, ul. Koszykowa 75, 00-662 Warsaw, Poland}
\date{\today}
\email{krzysztof.zberecki@pw.edu.pl}

\begin{abstract}

In this work we investigated adequacy of the Heisenberg model application to novel two-dimensional magnetic materials,
on an example of monolayer $\rm CrI_3$. We introduced the concept of the mean tensor invariant under symmetry operations
of the magnetic structure, which allows the number of parameters of the anisotropic tensor Heisenberg model to be significantly reduced, while maintaining the compliance with the results of ab-initio calculations. We derived the expressions for fourth-order corrections
to Heisenberg Hamiltonian and to Dzyaloshinskii–Moriya interaction in the form of quartic symmetry invariants with minimal number of parameters. We tested the physical adequacy of such approach in the case of monolayer $\rm CrI_3$, utilizing an alternative to four-states energy mapping --- the all-parameters least square fit. 

\end{abstract}

\pacs{*.*}

\maketitle

\section{Introduction}

Discovery of ferromagnetism in a single monolayer of $\rm CrI_3$ in 2017 \cite{Huang2017,Lin2018} raised the question of the possibility of using a simple effective Heisenberg spin model to describe ferromagnetic properties of two-dimensional (2D) materials. 
So far, ferromagnetism at finite temperatures in 2D materials has been experimentally observed, apart from $\rm CrI_3$, also in $\rm VSe_2$ \cite{1d} and $\rm MnSe_2$ \cite{2d}. 
Extensive theoretical and experimental research is carried out to predict the possibility of stable ferromagnetic and other magnetically ordered phases in 2D materials and to estimate their critical temperatures \cite{Torelli2019}. 
Mermin-Wagner theorem \cite{Mermin1966} predicts that 2D ferromagnetism is possible only when magnetic anisotropy is present, due to spin–orbit coupling.
The long–ranged order of magnetic moments in 2D materials are maintained at temperatures up to 45~K, as observed in $\rm CrI_3$ \cite{Huang2017}. 

Magnetic anisotropy in single layers of ferromagnetic materials usually favors the out-of-plane orientation of magnetic moments, such as in $\rm CrI_3$. The Hamiltonian describing single layers of $\rm CrI_3$ and other 2D ferromagnetic systems should contain terms describing exchange coupling and magnetic anisotropy. In materials having the inversion center between nearest-neighbors spin sites (e.g. Cr ions in $\rm CrI_3$ layers) anisotropy could not connected with the Dzyaloshinski-Moriya interaction (DMI), but may be related with anisotropic symmetric superexchange, and in $\rm CrI_3$ results mainly from the spin-orbit coupling on iodine atoms \cite{Xu2018}. 

Some of the 2D magnets were also analyzed in the first approximation using the Ising model of magnetism. However, this model is strictly correct only within the limit of the infinite single-ion anisotropy and provides only an upper bound for the critical temperature. Due to the fact that the local magnetic moments on Cr ions in $\rm CrI_3$ do not exhibit strong single ion anisotropy, a simple application of Ising model overestimates the Curie temperature for this material by a factor of three \cite{3d}. 

There is some controversy, which terms of magnetic interactions are most important in 2D materials. 
According to \cite{Lado2017}, single ion anisotropy parameter in $\rm CrI_3$ is much smaller than the exchange interaction constants between spins oriented along the axis perpendicular to the monolayer. Authors of \cite{Lado2017} also neglect non-diagonal elements of the tensor describing the exchange interaction, when one of its axes is oriented perpendicularly to the monolayer. To the contrary, the authors of \cite{Xu2018} indicate that these non-diagonal elements cannot be omitted, because they lead to the so-called Kitaev interactions.
 
To understand the mechanism of the exchange interaction and the sources of the magnetic anisotropy in 2D materials it is desirable to develop reliable effective models, and to provide
methods for the determination of parameters of their respective Hamitonians. 
The parameters of these models can be determined from ab-initio calculations performed for different magnetic configurations of 2D crystals with a few magnetic moments in the unit cell, then mapping ab-initio energies to a specific effective Heisenberg Hamiltonian. 
Such effective models would enable subsequent studies for larger systems and for finite temperatures, which are prohibitive by ab-initio calculations alone.

In the present paper, the number of the parameters in the effective Heisenberg Hamiltonian has been significantly decreased by analyzing the symmetry properties of the crystallographic spin structure described by the hexagonal lattice (such as in $\rm CrI_3$). For the 4th-order tensor in the Hamiltonian the number of parameters could be limited to 5 for both the symmetric and antisymmetric parts, and assuming more strict $SO(2)$ invariance condition, the number of parameters is reduced to only 3.
The values of parameters occurring in the symmetric part of Hamiltonian are determined for the single $\rm CrI_3$ monolayer by least-squares linear fit taking as input the results of 148 converged ab-initio calculations. The numerical procedure we applied yield low energy errors, therefore it can be used to reliably determine magnetic properties of 2D materials.

A recent alternative approach for construction of effective spin Hamiltonian models was proposed in \cite{Xue2020}, where instead of symmetry arguments machine learning techniques were applied to 2D MnO and multiferroic $\rm TbMnO_3$.

In sections \ref{secCH}-\ref{secF} of this paper, the effective Heisenberg Hamiltonian for two-dimensional crystals with hexagonal lattices is analyzed. By applying the symmetry properties of the system, formulas were derived connecting elements of 2nd and 4th order tensors related to the effective Hamiltonian with several parameters. In section \ref{secA} the results of ab-initio calculations performed for $\rm CrI_3$ monolayers are presented. Section \ref{secP} presents the mapping of the results of these calculation on the effective Hamiltonian together with the obtained estimation of its parameters for $\rm CrI_3$. The conclusions are set out in section \ref{secC}.   


\section{\label{secCH}Classical Heisenberg model}

Classical Heisenberg model from the phenomenological point of view corresponds to the lowest non-zero quadratic term
of the formal series expansion of a general magnetic Hamiltonian for a system composed of $n$ classical magnetic moments:

\begin{equation}
H\big(\vec{S}_1\,\ldots,\vec{S}_n\big)=E_0+\sum_{i,j=1}^N\vec{S}_i^{\,T}\cdot\hat{J}^{(ij)}\cdot\vec{S}_{j}+\ldots
\label{eq:ser}
\end{equation}

where

\begin{equation}
\vec{S}=\vec{\mu}/\mu
\label{eq:mu}
\end{equation}

is a dimensionless vector of magnetic moment reduced to unit length\footnote{Some authors prefer to multiply $\vec{S}$ by a numerical factor, e.g. when $\mu=3$ Bohr magnetons they multiply $\vec{S}$ by $\frac{3}{2}$, in analogy to quantum spin-$\frac{3}{2}$. This leads to trivial scaling of $\hat{J}$ matrix by $\frac{9}{4}$.},
which we will call "spin vector". $\hat{J}^{(ij)}$ are $ 3\times 3$ matrices of coefficients of interaction between spin vectors $\vec{S}_i$ and $\vec{S}_j$.

Due to the action of time-reversal operator $\hat{\cal{T}}$:

\begin{equation}
\hat{\cal{T}}(\vec{S})=-\vec{S}
\end{equation}

condition of time-reversal invariance requires that only even power terms should appear in the series (\ref{eq:ser}).
The lowest-order quadratic Hamiltonian for two spin interaction\footnote{Upper index (2) denotes two-spin interaction.}:

\begin{equation}
H^{(2)}\big(\vec{S}_1,\vec{S}_2\big)=\vec{S}_1^{\,T}\cdot\hat{J}\cdot\vec{S}_2
\label{eq:2body}
\end{equation}

depends on a single $3\times 3$ general matrix $\hat{J}$, which can always be decomposed into symmetric and antisymmetric parts:

\begin{equation}
\hat{J}=\hat{J}_S+\hat{J}_A
\end{equation}

where

\begin{equation}
\hat{J}_S=\frac{1}{2}\big(\hat{J}+\hat{J}^{\,T}\big)=\hat{J}_S^{\,T}\ ,\quad 
\hat{J}_A=\frac{1}{2}\big(\hat{J}-\hat{J}^{\,T}\big)=-\hat{J}_A^{\,T}
\end{equation}

The symmetric part

\begin{equation}
\hat{J}_S=\left(\begin{array}{rrr}
J_{11} & J_{12} & J_{13} \\
J_{12} & J_{22} & J_{23} \\
J_{13} & J_{23} & J_{33} \\
\end{array}\right)
\end{equation}

has 6 independent parameters, and the antisymmetric part

\begin{equation}
\hat{J}_A=\left(\begin{array}{rrr}
0 & D_3 & -D_2 \\
-D_3 & 0 & D_1 \\
D_2 & -D_1 & 0 \\
\end{array}\right)
\end{equation}

has 3 independent parameters and the resulting Hamiltonian is usually written in vectorial form of
Dzyaloshinskii–Moriya interaction (DMI):

\begin{equation}
H_{\rm DMI}(\vec{S}_1,\vec{S}_2)=\vec{S}_1^{\,T}\cdot\hat{J}_A\cdot\vec{S}_2=\vec{D}\cdot(\vec{S}_1\times\vec{S}_2)
\label{eq:DMI2}
\end{equation}

where $\vec{D}=(D_1,D_2,D_3)$ is a pseudo-vector, i.e. it should change sign when two spins are exchanged
in order to yield the Hamiltonian invariant under the permutation of spin vectors.

\begin{equation}
H_{\rm DMI}(\vec{S}_1,\vec{S}_2)=H_{\rm DMI}(\vec{S}_2,\vec{S}_1)
\end{equation}

\begin{figure}
\begin{center}
\includegraphics[width=0.5\textwidth]{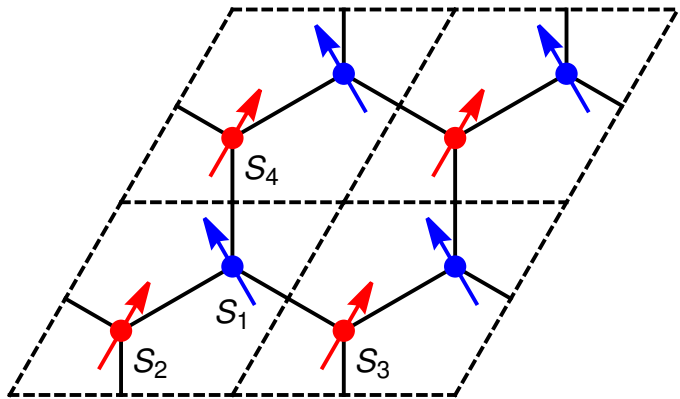}
\end{center}
\caption{Magnetic hexagonal lattice. Spin $\vec{S}_1$ with its three nearest neighbors: $\vec{S}_2$, $\vec{S}_3$, $\vec{S}_4$.}
\label{fig1}
\end{figure}

\section{Conditions of geometrical symmetry}

We consider two-dimensional magnetic crystals with atoms ordered in the hexagonal lattice.
Hamiltonian of the interaction of a given spin $\vec{S}_1$ with its three nearest neighbors $\vec{S}_2$, 
$\vec{S}_3$ and $\vec{S}_4$ (see Fig.~\ref{fig1}) should be invariant under the action of symmetry operator $\hat{R}$

\begin{equation}
H\big(\vec{S}_1,\vec{S}_2,\vec{S}_3,\vec{S}_4\big)=
H\big(\hat{R}\cdot\vec{S}_1,\hat{R}\cdot\vec{S}_4,\hat{R}\cdot\vec{S}_2,\hat{R}\cdot\vec{S}_3\big)
\label{eq:inv4}
\end{equation}

where $\hat{R}=R_z(2\pi/3)$ is the rotation around $z$-axis by $2\pi/3$, and $\hat{R}^3=\hat{1}$.
If we place the axis of rotation at atom with spin vector $\vec{S}_1$, rotation $\hat{R}$ not only act on spin vectors by also permutes spin-vectors at three neighboring atoms.
$\hat{R}$ is the generator of the group $C_3=\{\hat{1},\hat{R},\hat{R}^2\}$, which is the lowest symmetry group
for a two-dimensional crystal based on hexagonal lattice. Assuming that each of the two-spin interactions is described by
Eq.~(\ref{eq:2body}), in order to fulfill condition (\ref{eq:inv4}), we need to apply rotations $\hat{R}$ and $\hat{R}^2$
for spin pairs $\{\vec{S}_1,\vec{S}_3\}$ and $\{\vec{S}_1,\vec{S}_4\}$ respectively:

\begin{equation}
H(\vec{S}_1,\vec{S}_2,\vec{S}_3,\vec{S}_4\big)=H^{(2)}\big(\vec{S}_1,\vec{S}_2\big)+H^{(2)}\big(\hat{R}\cdot\vec{S}_1,\hat{R}\cdot\vec{S}_3\big)+
H^{(2)}\big(\hat{R}^2\cdot\vec{S}_1,\hat{R}^2\cdot\vec{S}_4\big)
\label{eq:R4}
\end{equation}

Eq. (\ref{eq:R4}) can be written as

\begin{align}
H(\vec{S}_1,\vec{S}_2,\vec{S}_3,\vec{S}_4\big)=\vec{S}_1^{\,T}\cdot\hat{J}\cdot\vec{S}_2+(\hat{R}\cdot\vec{S}_1)^{\,T}\cdot\hat{J}\cdot(\hat{R}\cdot\vec{S}_3)+
(\hat{R}^2\cdot\vec{S}_1)^{\,T}\cdot\hat{J}\cdot(\hat{R}^2\cdot\vec{S}_4)=
\nonumber
\\
\vec{S}_1^{\,T}\cdot\hat{J}\cdot\vec{S}_2+\vec{S}_1^{\,T}\cdot\hat{J}'\cdot\vec{S}_3+
\vec{S}_1^{\,T}\cdot\hat{J}''\cdot\vec{S}_4
\label{eq:JJ1}
\end{align}

Therefore, interaction of each pair of spin vectors is described by matrices of coefficients $\hat{J}'$ and $\hat{J}''$ which are related to matrix $\hat{J}$
by similarity transformation (rotation):

\begin{equation}
\hat{J}'=\hat{R}^{\,T}\cdot\hat{J}\cdot\hat{R}\ ,\quad \hat{J}''=\hat{R}^{\,T}\cdot\hat{J}'\cdot\hat{R}
\end{equation}

For an example of application of this method to magnetic $\rm CrI_3$ monolayer see \cite{Xu2018}.
One should note, that condition (\ref{eq:inv4}) of invariance of magnetic Hamiltonian for the system of spins $\{\vec{S}_1,\vec{S}_2,\vec{S}_3,\vec{S}_4\}$ 
leaves the number $M$ of independent parameters unchanged:
$M=6$ for symmetric quadratic exchange, $M=3$ for quadratic DMI interaction, $M=15$ for quartic (biquadratic) exchange.

In the phenomenological approach these parameters are not determined from first-principles, they merely serve as
arbitrary degrees of freedom for fitting a given Hamiltonian model to the results of ab-initio calculations of magnetic energy $E(\vec{S}_1,\ldots,\vec{S}_n)$. Therefore, for a trustworthy and physically meaningful phenomenological model the less the number of independent parameters is the better.

The number of independent parameters can be significantly lowered by assuming more strict invariance condition than (\ref{eq:inv4}):

\begin{equation}
H^{(2)}(\vec{S}_1,\vec{S}_2)=H^{(2)}(\hat{R}\cdot\vec{S}_1,\hat{R}\cdot\vec{S}_2)
\label{eq:inv}
\end{equation}

This condition of invariance of two spin interaction formally follows from (\ref{eq:inv4}) when

\begin{equation}
\hat{S}_2=\hat{S}_3=\hat{S}_4
\label{eq:S3}
\end{equation}

then from Eq.~(\ref{eq:JJ1})

\begin{equation}
H(\vec{S}_1,\vec{S}_2=\vec{S}_3=\vec{S}_4\big)=3\,\vec{S}_1^{\,T}\cdot\hat{J_m}\cdot\vec{S}_2
\end{equation}

where

\begin{equation}
\hat{J}_m=\frac{1}{3}\big(\hat{J}+\hat{J}'+\hat{J}''\big)
\label{eq:Jm}
\end{equation}

The mean tensor $\hat{J}_m$ is invariant under rotations

\begin{equation}
\hat{R}^{\,T}\cdot\hat{J}_m\cdot\hat{R}=\hat{J}_m
\label{eq:invR}
\end{equation}

The condition (\ref{eq:inv}) ensures the minimal number of parameters for quadratic and quartic magnetic Hamiltonians.
Eq.~(\ref{eq:S3}) means that the magnetic structure is reduced to the single unit cell, i.e. spin vectors in all unit cells should be the same, as depicted in Fig.~\ref{fig1}. For practical applications instead of supercell approach, e.g. utilized
for two-dimensional magnetic crystals in \cite{Torelli2019}, one should evaluate ab-initio magnetic configurations with arbitrary angles between spin vectors in a single unit cell.

\section{Second-order invariant Hamiltonians}

Symmetric tensor $J_{ij}=J_{ji}$ invariant under $R_z(2\pi/3)$ has only two independent components.
Hamiltonian built from this tensor\footnote{We use the Einstein summation convention for Cartesian indices}:

\begin{equation}
H^{(2)}(\vec{S}_1,\vec{S}_2)=J_{ij}\,S_{1i}\,S_{2j}=J_1\,\big(S_{1x}S_{2x}+S_{1y}S_{2y}\big)+J_2\,S_{1z}S_{2z}
\label{eq:hs2}
\end{equation}

describes the well-known anisotropic exchange interaction,
where in the following we will utilize simplified single index notation for independent components,
in this case $J_{ij}=J_{k}$ with single consecutive index $k=1,2$.

Validity of Eq.~(\ref{eq:hs2}) derived from the condition of $SO(2)$-invariance for two dimensional magnetic crystals can be verified e.g. in the case of magnetic $\rm CrI_3$ monolayer, by
taking six components of general symmetric tensor $\hat{J}$ determined in \cite{Xu2018} and performing matrix average (\ref{eq:Jm}).
As the result one obtains diagonal tensor with only two independent components $J_1=-2.11\ \mbox{meV}$ and $J_2=-2.23\ \mbox{meV}$.
If we apply the same method to the general symmetric tensor $\hat{J}$ determined in \cite{Sabani2020} we obtain
diagonal tensor with two independent components $J_1=-4.46\ \mbox{meV}$ and $J_2=-4.63\ \mbox{meV}$. The discrepancy between the results of these
two papers is due to the fact that in \cite{Xu2018} there appears a factor $\frac{1}{2}$ in the definition of Heisenberg Hamiltonian. From the condition $J_1\approx J_2$ we infer that the magnetic energy of $\rm CrI_3$  monolayer is weakly anisotropic.

Antisymmetric tensor $J_{ij}=-J_{ji}$ invariant under $R_z(2\pi/3)$ has only one independent component.
Hamiltonian built from this tensor:

\begin{equation}
H_{\rm DMI}(\vec{S}_1,\vec{S}_2)=J_{ij}\,S_{1i}\,S_{2j}=J_1\,\big(S_{1x}S_{2y}-S_{2x}S_{1y}\big)
\label{eq:ha2}
\end{equation}

describes the DMI interaction (\ref{eq:DMI2}) with single non-zero out-of plane component of interaction vector 
$\vec{D}=(0,0,J_1)$. DMI interaction may exist only when the inversion symmetry is broken, e.g. for magnetic $\rm CrI_3$
monolayer with applied external electric field \cite{Liu2018}. Note that in paper \cite{Liu2018} a non-zero in-plane component $D_{xy}$ of DMI interaction is discussed, but only for a single pair of spin-vectors. For the mean tensor $\hat{J}_m$ in  Eq.~(\ref{eq:Jm}) all in-plane contributions to DMI interaction cancel, and the total energy of the system depends only on the out-of-plane DMI component~$D_z$

Magnetocrystalline anisotropy (MCA) based on symmetric tensor $A_{ij}=A_{ji}$ invariant under $R_z(2\pi/3)$ has only two independent components. Hamiltonian built from this tensor:

\begin{equation}
H_{\rm MCA}(\vec{S})=A_{ij}\,S_{i}\,S_{j}=A_1\,\big(S_{x}^2+S_{y}^2\big)+A_2\,S_{z}^2
\label{eq:hma2}
\end{equation}

can be reduced to simpler form using the condition 

\begin{equation}
S^2=S_x^2+S_y^2+S_z^2=\mbox{const}
\label{eq:S2}
\end{equation}

and setting $A=A_2-A_1$ for a single anisotropy constant:

\begin{equation}
H_{\rm MCA}=A\,S_{z}^2+\mbox{const}
\label{eq:mca2}
\end{equation}

\section{\label{secF}Fourth-order invariant Hamiltonian}

Fourth-order Hamiltonian for interaction of two spins $\vec{S}_1$ and $\vec{S}_2$ built from 4th-order general tensor $K_{ijkl}$, has the following form

\begin{equation}
H^{(2)}(\vec{S}_1,\vec{S}_2)=K_{ijkl}\,S_{1i}\,S_{1j}\,S_{2k}\,S_{2l}
\label{eq:sum4}
\end{equation}

If we assume that magnetic interaction is invariant under permutation of spin vectors:

\begin{equation}
H^{(2)}(\vec{S}_1,\vec{S}_2)=H^{(2)}(\vec{S}_2,\vec{S}_1)
\end{equation}

and noting that permutation of individual components of spin $\vec{S}_1$ or $\vec{S}_2$ does not change the sum in equation (\ref{eq:sum4}) we obtain the condition of full-permutational invariance:

\begin{equation}
K_{ijkl}=K_{(ijkl)}
\end{equation}

where $(ijkl)$ is any permutation of four Cartesian indices. Full-permutational invariant 4th-order tensor has 15 independent components.

We can further reduce the number of independent components (parameters of the Hamiltonian) by applying condition (\ref{eq:inv}) of invariance under symmetry operator $\hat{R}$:

\begin{equation}
H^{(2)}(\vec{S}_1,\vec{S}_2)=H^{(2)}(\hat{R}\cdot\vec{S}_1,\hat{R}\cdot\vec{S}_2)=
K_{ijkl}\,\big(\hat{R}\cdot \vec{S}_1\big)_i\,
\big(\hat{R}\cdot \vec{S}_1\big)_j\,
\big(\hat{R}\cdot \vec{S}_2\big)_k\,
\big(\hat{R}\cdot \vec{S}_2\big)_l\,
\end{equation}

which, valid for any spin vectors $\vec{S}_1$ and $\vec{S}_2$, leads to the invariance condition for the $K_{ijkl}$ tensor itself

\begin{equation}
K_{i'j'k'l'}=K_{ijkl}\,R_{ii'}\,R_{jj'}\,R_{kk'}\,R_{ll'}
\end{equation}

where $R_{ij}$ is an orthogonal $3\times 3$ matrix representing the action of symmetry operator $\hat{R}$ in three-dimensional space.

Application of $C_3$ symmetry, i.e. the lowest symmetry for a hexagonal lattice, results in Hamiltonian (\ref{eq:sum4}) which has only five parameters $K_1,\ldots,K_5$. This Hamiltonian can be represented in the following tabular form:

\begin{equation}
\begin{array}{|l|l|}
\hline
 K_1 & \frac{4}{3} S_{1x}S_{2x}S_{1y}S_{2y}+S^2_{1x}S^2_{2x}+S^2_{1y}S^2_{2y}+
 \frac{1}{3}\big(S^2_{1x}S^2_{2y}+S^2_{2x}S^2_{1y}\big) \\
 \hline
 K_2 & 2S_{1z}\big[S_{1x}\big(S^2_{2x}-S^2_{2y}\big)-2S_{2x}S_{1y}S_{2y}\big]+
 2S_{2z}\big[S_{2x}\big(S^2_{1x}-S^2_{1y}\big)-2S_{1x}S_{2y}S_{1y}\big]
 \\
 \hline
 K_3 & 2S_{1z}\big[S_{1y}\big(S^2_{2y}-S^2_{2x}\big)-2S_{2y}S_{1x}S_{2x}\big]+
 2S_{2z}\big[S_{2y}\big(S^2_{1y}-S^2_{1x}\big)-2S_{1y}S_{2x}S_{1x}\big] \\
 \hline
 K_4 & \big(S^2_{1x}+S^2_{1y}\big)S^2_{2z}+\big(S^2_{2x}+S^2_{2y}\big)S^2_{1z}+
 4\big(S_{1x}S_{2x}+S_{1y}S_{2y}\big)S_{1z}S_{2z} \\
 \hline
 K_5 & S^2_{1z} S^2_{2z} \\
 \hline
\end{array}
\label{eq:hc3}
\end{equation}

Each of the five terms of this Hamiltonian is invariant under action of symmetry operator $R_z(2\pi/3)$, and does not change under the permutation of spin vectors $\vec{S}_1\leftrightarrow \vec{S}_2$.

We can further reduce number of parameters of 4th-order Hamiltonian (\ref{eq:sum4}) by assuming more strict invariance under group $SO(2)$ of all rotations around $z$-axis by any angle $\alpha$. Discrete group $C_3$ describes the lowest symmetry of an atomic hexagonal lattice.  The continuous symmetry of a two-dimensional plane perpendicular to the $z$-axis neglects the details of any atomic structure of the magnetic crystal. This assumption is analogous to that of the $SO(3)$ group invariance of Heisenberg Hamiltonian for bulk crystals, from which follows that the magnetic interaction of two spins $\vec{S}_1$ and $\vec{S}_2$ should be proportional to even powers of the scalar product $\vec{S}_1\cdot\vec{S}_2$, which is invariant under any rotations in three-dimensional space \cite{Xue2020}.

Application of $SO(2)$-invariance to 2nd-order Hamiltonians (\ref{eq:hs2}), (\ref{eq:ha2}) and (\ref{eq:hma2}) leaves them unchanged, with respect to their $C_3$-invariant forms. However,
the requirement of $SO(2)$-invariance applied to 4th-order Hamiltonian (\ref{eq:sum4}) further reduces the number of its independent parameters to only three.
4th-order Hamiltonian can thus be simplified to the linear combination of three quartic $SO(2)$ invariants:

\begin{equation}
\begin{array}{|l|l|}
\hline
 K_1 & 4 S_{1x}S_{2x}S_{1y}S_{2y}+3\big(S^2_{1x}S^2_{2x}+S^2_{1y}S^2_{2y}\big)+
 S^2_{1x}S^2_{2y}+S^2_{2x}S^2_{1y} \\
 \hline
 K_2 & \big(S^2_{1x}+S^2_{1y}\big)S^2_{2z}+\big(S^2_{2x}+S^2_{2y}\big)S^2_{1z}+
 4\big(S_{1x}S_{2x}+S_{1y}S_{2y}\big)S_{1z}S_{2z} \\
 \hline
 K_3 & S^2_{1z} S^2_{2z} \\
 \hline
\end{array}
\label{eq:hso2}
\end{equation}

and may be obtained from $C_3$-invariant Hamiltonian (\ref{eq:hc3}) by simply setting $K_2=K_3=0$, multiplying $K_1$ by 3, and renumbering the remaining parameters.

Let us consider now the extension of Dzyaloshinskii–Moriya interaction to 4th-order terms based on general expression (\ref{eq:sum4}). Tensor $K_{ijkl}$ should still be symmetric under permutation of individual components of spin $\vec{S}_1$ or $\vec{S}_2$:

\begin{equation}
K_{ijkl}=K_{jikl}\ \quad\mbox{and}\quad
K_{ijkl}=K_{ijlk}
\label{eq:c1}
\end{equation}

However, now we assume that it is antisymmetric under permutation of spin vectors $\vec{S}_1\leftrightarrow \vec{S}_2$.

\begin{equation}
K_{ijkl}=-K_{klij}
\label{eq:c2}
\end{equation}

Tensor $K_{ijkl}$ fulfilling conditions (\ref{eq:c1}) and (\ref{eq:c2}) has 15 independent components.
If we add invariance condition (\ref{eq:inv}) for symmetry operation $R_z(2\pi/3)$ ($C_3$-invariance) we obtain the following Hamiltonian with five independent parameters:

\begin{equation}
\begin{array}{|l|l|}
\hline
 K_1 & S^2_{2z}\big(S^2_{1x}+S^2_{1y}\big)-S^2_{1z}\big(S^2_{2x}+S^2_{2y}\big) \\
 \hline
 K_2 & 2\big(S_{2x}S_{1y}-S_{1x}S_{2y}\big)\big(S_{1x}S_{2x}+S_{1y}S_{2y}\big)
 \\
 \hline
 K_3 & 
 2S_{1x}S_{1z}\big(S^2_{2x}-S^2_{2y}\big)
 - 2S_{2x}S_{2z}\big(S^2_{1x}-S^2_{1y}\big)
+ 4S_{1y}S_{2y}\big(S_{1x}S_{2z}-S_{2x}S_{1z}\big) \\
 \hline
 K_4 & 
 2S_{1y}S_{1z}\big(S^2_{2x}-S^2_{2y}\big)
 -2S_{2y}S_{2z}\big(S^2_{1x}-S^2_{1y}\big)
 +4S_{1x}S_{2x}\big(S_{2y}S_{1z}-S_{1y}S_{2z}\big) \\
 \hline
 K_5 & 4S_{1z} S_{2z}\big(S_{2x}S_{1y}-S_{1x}S_{2y}\big) \\
 \hline
\end{array}
\label{eq:hc31}
\end{equation}

Each term of this Hamiltonian is invariant under rotation by $2\pi/3$ around $z$-axis and changes sign
under permutation of spin vectors.

More strict condition of $SO(2)$-invariance reduces the number of independent parameters of 4th-order DMI Hamiltonian to only three.
The resulting Hamiltonian which consists of a linear combination of three quartic terms has the following form:

\begin{equation}
\begin{array}{|l|l|}
\hline
 K_1 & S^2_{2z}\big(S^2_{1x}+S^2_{1y}\big)-S^2_{1z}\big(S^2_{2x}+S^2_{2y}\big) \\
 \hline
 K_2 & 2\big(S_{2x}S_{1y}-S_{1x}S_{2y}\big)\big(S_{1x}S_{2x}+S_{1y}S_{2y}\big)
 \\
 \hline
 K_3 & 4S_{1z} S_{2z}\big(S_{2x}S_{1y}-S_{1x}S_{2y}\big) \\
 \hline
\end{array}
\label{eq:DMI4}
\end{equation}

and may be obtained from (\ref{eq:hc31}) by setting $K_3=K_4=0$ and renaming $K_5$ to $K_3$.
Hamiltonian (\ref{eq:DMI4}) may be written in the coordinate-independent vectorial form:

\begin{equation}
H_{\rm DMI}(\vec{S}_1,\vec{S}_2)=K_1\big[(\vec{S}_2\cdot\vec{S}_2)_\perp(\vec{S}_1\cdot\vec{S}_1)_{||}
-(\vec{S}_1\cdot\vec{S}_1)_\perp(\vec{S}_2\cdot\vec{S}_2)_{||}\big]
+(\vec{S}_1\times\vec{S}_2)_\perp\big[K_2(\vec{S}_1\cdot\vec{S}_2)_{||}+
K_3(\vec{S}_1\cdot\vec{S}_2)_\perp
\big]
\label{eq:DMI4a}
\end{equation}

by introduction of in-plane $||$ and out-of plane $\perp$ parts of scalar and vector products:

\begin{equation}
(\vec{S}_i\cdot\vec{S}_j)_\perp=S_{iz}S_{jz}\ ,\quad
(\vec{S}_i\cdot\vec{S}_j)_{||}=S_{ix}S_{jx}+S_{iy}S_{jy}\ ,\quad
(\vec{S}_i\times\vec{S}_j)_\perp=S_{ix}S_{jy}-S_{iy}S_{jx}
\end{equation}

and by redefinition of Hamiltonian parameters:  $K_2\rightarrow -2K_3$, $K_3\rightarrow -4K_3$.
Eq.~(\ref{eq:DMI4a}) constitutes the 4th-order generalization of expression (\ref{eq:DMI2}) for DMI interaction applicable to two-dimensional magnetic crystals.

The 4th-order correction to magnetocrystalline anisotropy can be built from fully symmetric 4th-order tensor 
$B_{ijkl}=B_{(ijkl)}$
in an analogy to Eq.~(\ref{eq:sum4})

\begin{equation}
H_{\rm MCA}(\vec{S})=B_{ijkl}S_iS_jS_kS_l
\end{equation}

Application of $C_3$-invariance condition $H_{\rm MCA}(\vec{S})=H_{\rm MCA}(\hat{R}\cdot\vec{S})$, where $\hat{R}=R_z(2\pi/3)$ leads to the following
Hamiltonian with five independent parameters:

\begin{equation}
H_{\rm MCA}(\vec{S})=B_1\big(S^2_x+S^2_y\big)^2+4B_2S_xS_z\big(S^2_x-3S^2_y\big)+
4B_3S_yS_z\big(S^2_y-3S^2_x\big)+6B_4S^2_z\big(S^2_x+S^2_y)+B_5S^4_z
\label{eq:ma4}
\end{equation}

Utilizing the condition (\ref{eq:S2}), omitting constant terms, and disregarding terms quadratic in spin components, Eq.~(\ref{eq:ma4}) can be simplified to a three-parameter form

\begin{equation}
H_{\rm MCA}(\vec{S})=BS_z^4+4S_z\big[B_2S_x\big(S_x^2-3S_y^2\big)+B_3S_y\big(S_y^2-3S_x^2\big)\big]
\label{eq:mca4}
\end{equation}

where $B=B_1-6B_4+B_5$. 

$SO(2)$-invariant 4th-order magnetocrystalline anisotropy after the same manipulations could be reduced to much simpler
form containing only one quartic term $BS_z^4$.


\section{\label{secA}Ab-initio calculations}

\begin{figure}[ht]
  \begin{center}
      \includegraphics[width=0.5\textwidth]{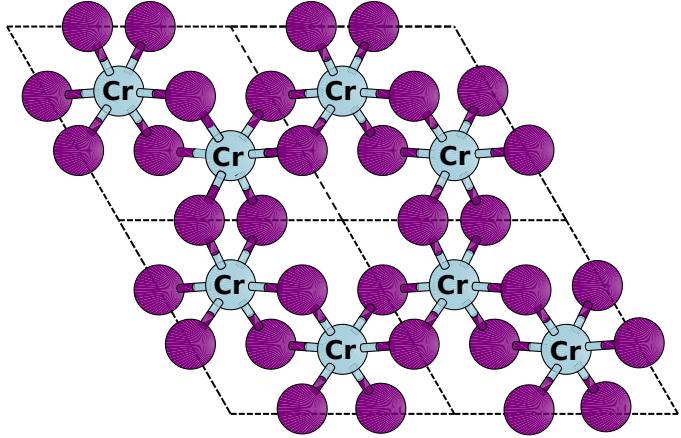}
  \end{center}
    \vspace{-1em}
    \caption{Four unit cells of $\rm CrI_3$ monolayer, light blue balls -- Cr atoms, dark violet balls -- I atoms.}
    \label{fig1a}
\end{figure}

Recently, $\rm CrI_3$ has become a canonical system for exploring magnetism in 2D, since it was one of the very first 2D magnets discovered, with a decent Curie temperature equal to 45 K \cite{Huang2017}. In $\rm CrI_3$, the large spin-orbit coupling (SOC) in the I ions creates a considerable magnetic anisotropy \cite{Lado2017}, which makes the breaking of the Mermin-Wagner theorem~\cite{Mermin1966}  possible.
$\rm CrI_3$ monolayer crystallizes in a hexagonal lattice with P-31m symmetry group. There are eight atoms in the unit cell, two chromium atoms and six iodine atoms, as shown in Fig.~\ref{fig1a}.

According to the previous ab-initio calculations \cite{Lado2017,Zhang2015,Liu2016,Torelli2019,Sadhukhan2022}, $\rm CrI_3$ is an FM semiconductor with a magnetic moment equal to 6.0 $\mu_{\rm B}$ per unit cell, almost entirely residing on the Cr atoms. This is in agreement with the experiment, where the 3.0 $\mu_{\rm B}$ per Cr atom was observed for the bulk $\rm CrI_3$ \cite{McGuire2015}. 

Magnetic anisotropy energy (MAE) is defined as the difference between the total energies corresponding to the magnetization in the off-plane and in-plane directions 

\vspace{-1em}
\begin{equation}
E_{\rm MAE}=E_\parallel-E_\perp
\label{eq:MAE}
\end{equation}

According to \cite{Torelli2019}, for monolayer $\rm CrI_3$ $E_{\rm MAE}$ equals $1.71\ \mbox{meV}$, indicating the off-plane easy axis. 

\subsection{Calculation details}

To test the versatility of our model, we conducted a series of ab-initio calculations for the $\rm CrI_3$ monolayer. We used the Density Functional Theory (DFT), as implemented in the VASP code \cite{Kresse1993,Kresse1994,Kresse1996,Kresse1996a}, with projector augmented wave (PAW) pseudopotentials \cite{Blochl1994,Kresse1999} and PBE \cite{Perdew1996} (GGA) functional. For the sampling of the Brillouin zone, a dense $20\times 20 \times 1$ grid was used, while the plane wave energy cutoff was set to 500 eV. All the structures were optimized until the forces exerted on atoms were smaller than $10^{-5}\ \mbox{eV}/\angstr$. In all our calculations, the SOC was included, as implemented in the PAW formalism by Steiner et. al. \cite{Steiner1993}. To include on-site repulsion in $d$-shell electrons in Cr atoms, we used the GGA+U scheme in the rotationally invariant approach introduced by Liechtenstein et. al. \cite{Lichtenstein1995}. The value of $U$ was taken equal to 3.5 eV, as determined in \cite{Wang2006}.

\subsection{Calculation results}

The magnetic ground state of CrI$_{3}$ is the ferromagnetic (FM) one, while the antiferromagnetic (AFM) configuration exhibits energy higher by 42.32 meV (for $U=0\ \mbox{eV}$). The MAE was determined as equal to $-1.48\ \mbox{meV}$ for $U=0\ \mbox{eV}$ and 
$-1.70\ \mbox{meV}$ for $U=3.5\ \mbox{eV}$. This is in agreement with \cite{Torelli2019}, where was shown, that addition of GGA+U scheme lowers the magnetic energy.

Applicability of the Heisenberg model is based on energy dependence on angles between spins in the form of a continuous $(\theta,\varphi)$ dependence, so to fit the model, one needs a dense grid of data. We prepared 250 starting configurations, with the direction of magnetic moments on Cr atoms distributed randomly on the unit sphere. For $U=0\ \mbox{eV}$, we obtained 245 converged cases, while for $U=3.5\ \mbox{eV}$ we got 148 cases.

\begin{figure}[ht]
  \begin{center}
      \includegraphics[width=0.4\textwidth]{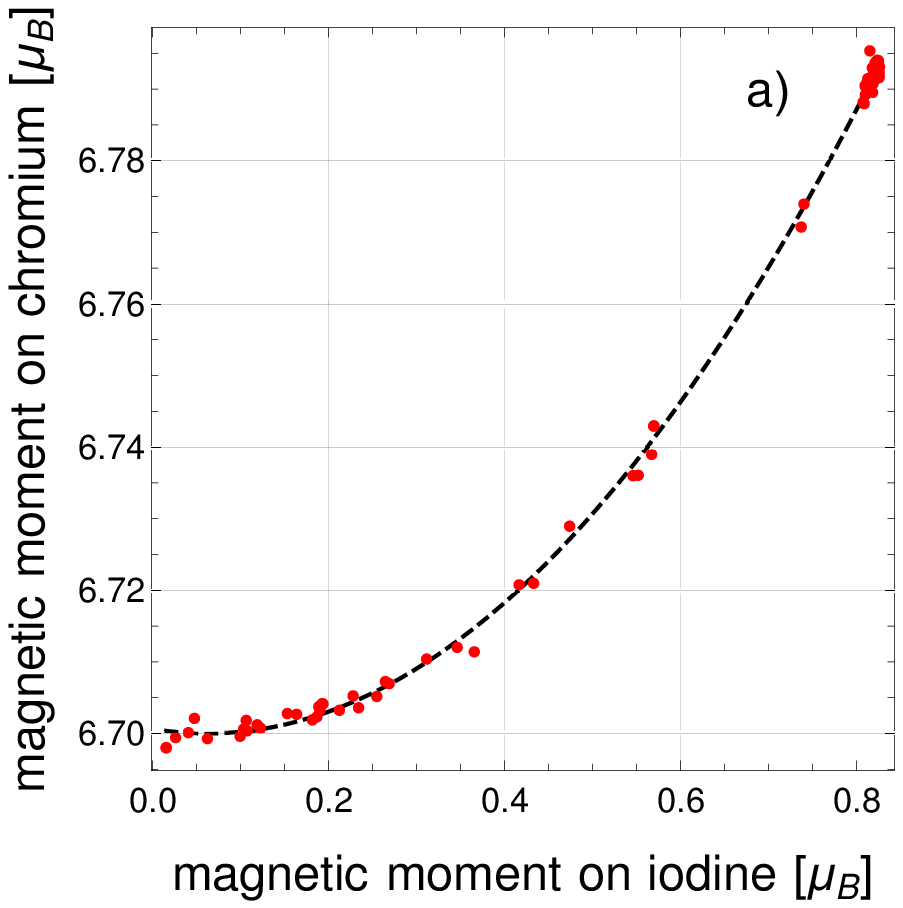}
      \includegraphics[width=0.42\textwidth]{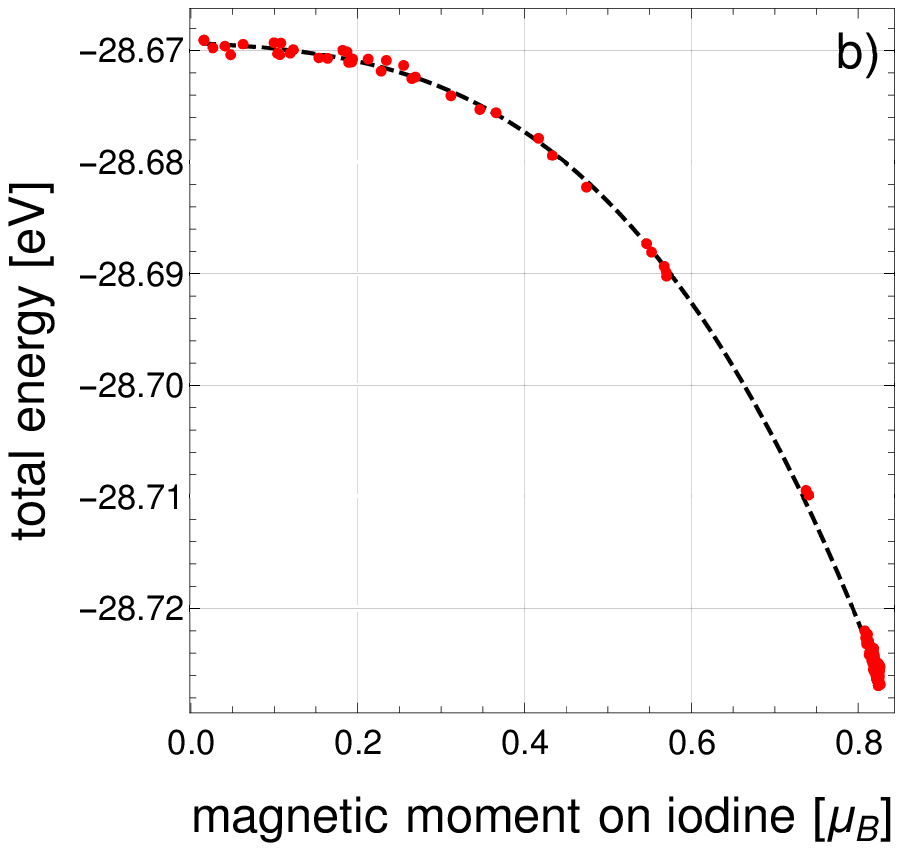}
      \end{center}
    \vspace{-1em}
    \caption{a) The total magnetic moment on iodine atoms per unit cell vs the total moment on chromium atoms. Dashed line represents quadratic fit.
    b) The total magnetic moment on iodine atoms per unit cell vs the total energy of the system. Dashed line represents polynomial fit. Results of ab-initio calculations for $U=3.5\ \mbox{eV}$.}
    \label{fig1b}
\end{figure}

Despite setting the initial values of magnetic moments on I atoms as zero, non-zero values of these moments are obtained in the final converged magnetic configurations. Magnetic moments on iodine are always directed anti-parallel to the magnetic moments on the Cr atoms. The total magnetic moment on iodine atoms per unit cell vs the total moment on chromium atoms is presented in Fig. \ref{fig1b}, obtained for $U=3.5\ \mbox{eV}$.

\begin{figure}[ht]
  \begin{center}
      \includegraphics[width=0.4\textwidth]{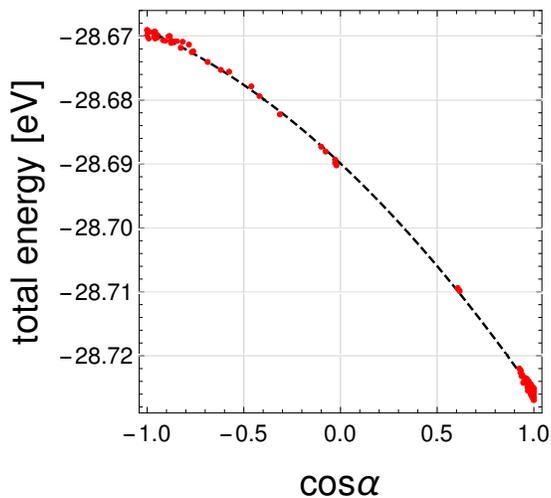}
  \end{center}
    \vspace{-0.5em}
    \caption{Dependence of the $\alpha$ angle (cosine of $\alpha$) between the magnetic moments on the Cr atoms in a unit cell on the total energy of the configuration. $\cos\alpha=1$ ($\alpha=0$) corresponds to FM configuration and $\cos\alpha=-1$ ($\alpha=180^\circ$) corresponds to AFM configuration. Dashed line represents quadratic fit. Results of ab-initio calculations for $U=3.5\ \mbox{eV}$.}
    \label{fig1c}
\end{figure}

As can be seen, this relation is parabolic, leading to some fluctuations of the total moment. The magnetic moments on I concentrate mainly in two ranges of values: [0,0.2] $\mu_{B}$ and [0.75,0.85] $\mu_{B}$. Therefore, the total magnetic moment on I can be treated as an order parameter to distinguish magnetic FM and AFM structures separated by the value of $\mu_{\rm I}=0.5\ \mu_{B}$. It is visible as well in the plot (Fig. \ref{fig1c}) showing the dependence of the $\cos\alpha$, where $\alpha$ is the angle between the magnetic moments on two Cr atoms in the unit cell, on the total energy $E$ of the configuration ($\cos\alpha=1$ corresponds to FM configuration and $\cos\alpha=-1$ corresponds to AFM configuration). For spin vectors  $\vec{S}_1$ and $\vec{S}_2$ as in Eq.~(\ref{eq:mu}) 
$\cos\alpha=\vec{S}_1\cdot\vec{S}_2$.
Parabolic dependence $E(\cos\alpha)$ suggests a significant role of quartic corrections to the magnetic energy of $\rm CrI_3$, proportional to $\cos^2\alpha=(\vec{S}_1\cdot\vec{S}_2)^2$.

\begin{figure}[ht]
  \begin{center}
   \includegraphics[width=0.4\textwidth]{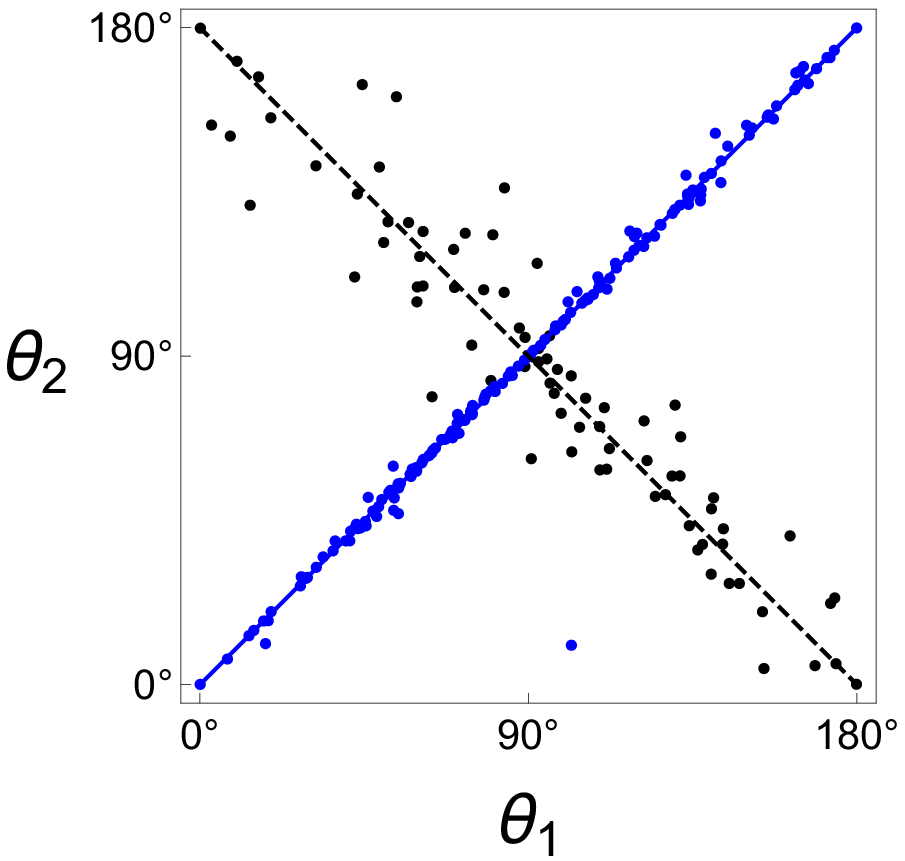}
   \includegraphics[width=0.4\textwidth]{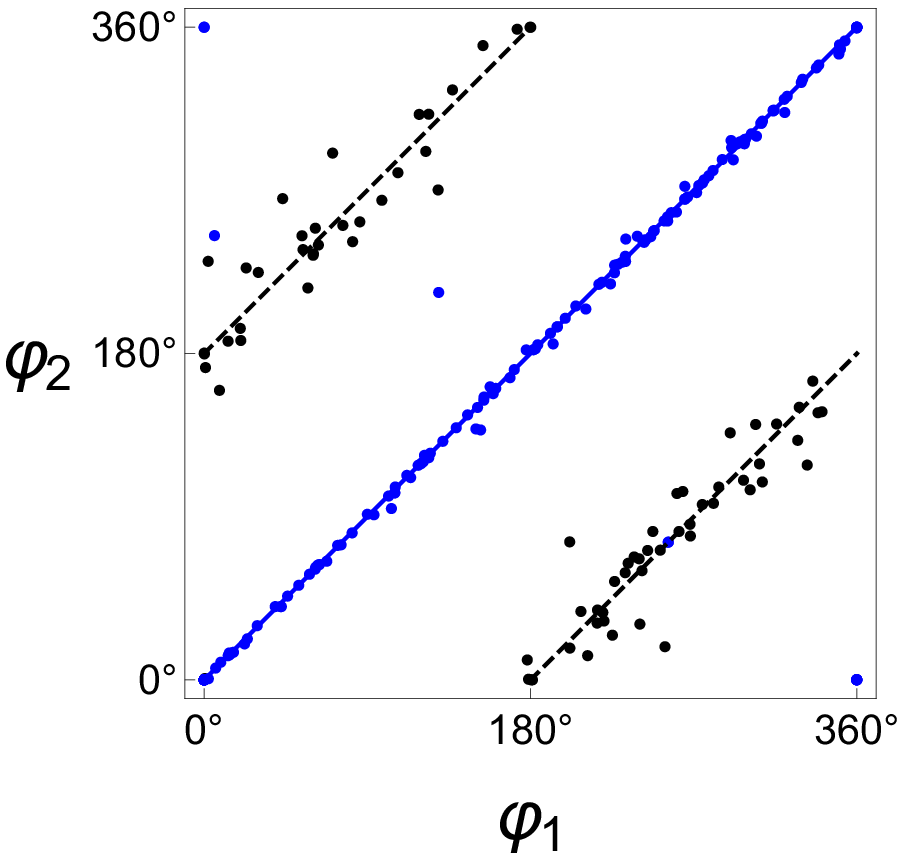}
    \vspace{-1em}
    \end{center}
    \caption{Spherical angles $\theta_1$ and $\theta_2$ (left panel) and $\varphi_1$ and $\varphi_2$ (right panel) for ab-initio magnetic configurations for Hubbard $U=0$. Black dots symbolize cases with total magnetic moment on iodine atoms less than 0.2 $\mu_{\rm B}$. Blue dots symbolize cases with total magnetic moment on iodine atoms greater than 0.2 $\mu_{\rm B}$. \newline
    In left panel: black dashed line represents AFM orientation: $\vec{S}_2=-\vec{S}_1$, $\theta_2=180^\circ-\theta_1$, 
blue solid line represents FM orientation: $\vec{S}_2=\vec{S}_1$, $\theta_2=\theta_1$.
\newline
    In right panel: black dashed line represents AFM orientation: $\vec{S}_2=-\vec{S}_1$, $\varphi_2=180^\circ+\varphi_1$,
blue solid line represents FM orientation: $\vec{S}_2=\vec{S}_1$, $\varphi_2=\varphi_1$.}    
    \label{fig2}
\end{figure}

\begin{figure}[ht]
  \begin{center}
      \includegraphics[width=0.4\textwidth]{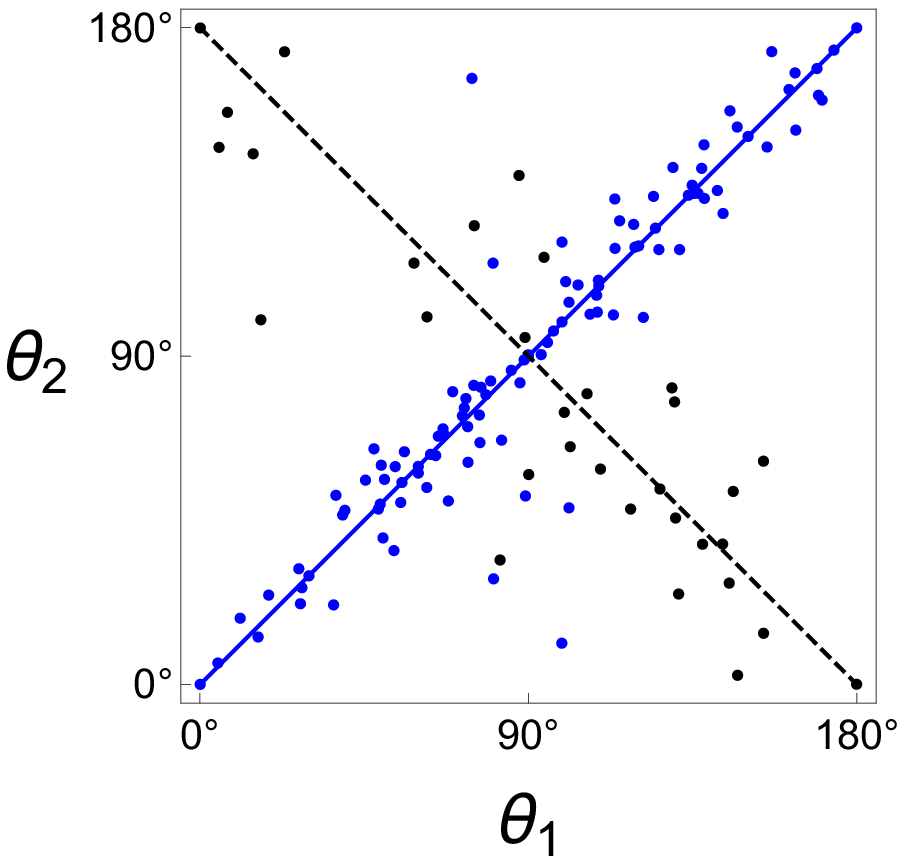}
      \includegraphics[width=0.4\textwidth]{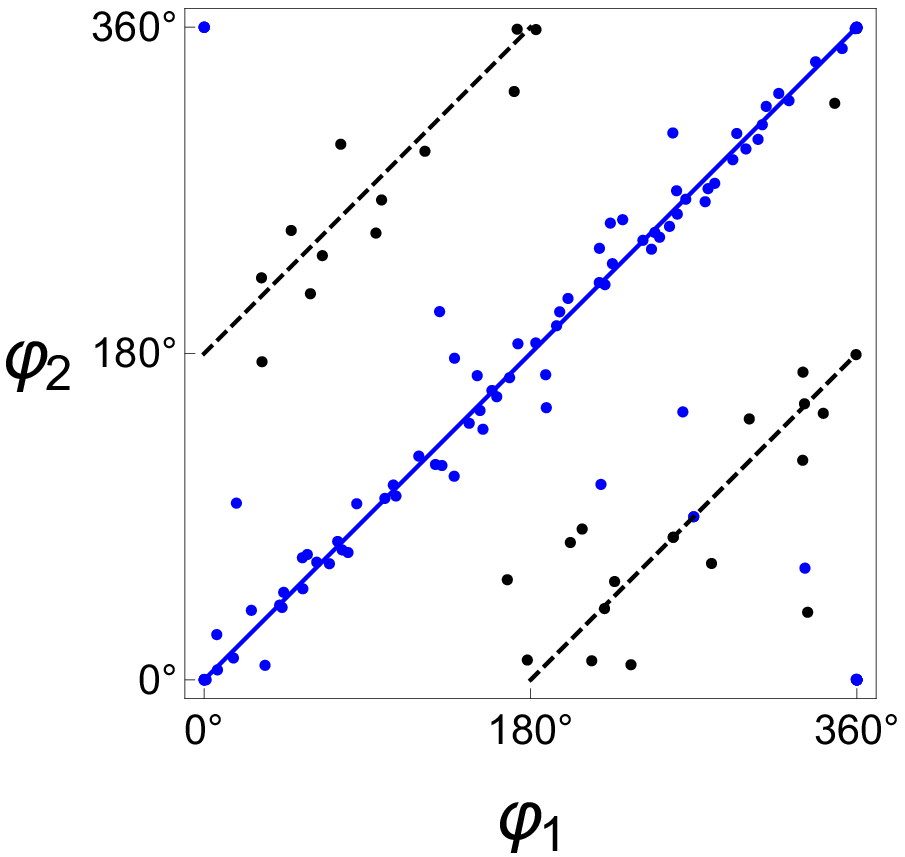}
        \end{center}
    \vspace{-1em}
    \caption{Spherical angles $\theta_{1}$ and $\theta_{2}$ (left panel) and $\phi_{1}$ and $\phi_{2}$ (right panel) for ab-initio magnetic configurations for Hubbard $U=3.5$. Black dots symbolize cases with total magnetic moment on iodine atoms less than 0.5 $\mu_{\rm B}$. Blue dots symbolize cases with total magnetic moment on iodine atoms greater than 0.5 $\mu_{\rm B}$. 
\newline
    In left panel: black dashed line represents AFM orientation: $\vec{S}_2=-\vec{S}_1$, $\theta_2=180^\circ-\theta_1$, 
blue solid line represents FM orientation: $\vec{S}_2=\vec{S}_1$, $\theta_2=\theta_1$.
\newline
    In right panel: black dashed line represents AFM orientation: $\vec{S}_2=-\vec{S}_1$, $\varphi_2=180^\circ+\varphi_1$,
blue solid line represents FM orientation: $\vec{S}_2=\vec{S}_1$, $\varphi_2=\varphi_1$.}    
    \label{fig3}
\end{figure}

Figs. \ref{fig2} and \ref{fig3} depict the distribution of spherical angles for obtained magnetic moments for $U=3.5\ \mbox{eV}$, belonging to two mentioned families. In the case of configurations with moments on iodine atoms greater than 0.5 $\mu_{B}$, maximum single magnetic moment on I atoms is equal to $0.11\ \mu_{\rm B}$, while it is $3.28\ \mu_{\rm B}$ on single Cr atom, so the assumption that the total moment is localized on Cr atoms remains valid.

\section{\label{secP}Parameters of Heisenberg models for $\textbf{CrI}_{\textbf 3}$}

\subsection{Isotropic Hamiltonian and magnetic anisotropy}

Quadratic dependence of total energy on cosine of angle between spin vectors $\vec{S}_1$ and $\vec{S}_2$ on Cr atoms,
as presented in Fig.~\ref{fig1c},
can be described by isotropic biquadratic Heisenberg model

\begin{equation}
H(\vec{S}_1,\vec{S}_2)=E_0+3\big[J\big(\vec{S}_1\cdot\vec{S}_2\big)+K\big(\vec{S}_1\cdot\vec{S}_2\big)^2\big]
\label{eq:hiso}
\end{equation}

which depends only on $\cos\alpha=\vec{S}_1\cdot\vec{S}_2$, with two parameters $J$ and $K$. Factor 3 takes into account interaction
of a given magnetic moment with three nearest neighbors.
From the parabolic least-square fit (see Fig.~\ref{fig1c}) we obtain $J=-9.4\ \mbox{meV}$ and $K=-2.5\ \mbox{meV}$.
The value of $J$ agrees with the results of \cite{Xu2018,Sabani2020} after application of tensor average (\ref{eq:Jm})
if we take into account that in these papers $S=\frac{3}{2}$, and we take $S=1$, hence the additional $\frac{9}{4}$ factor
in Heisenberg Hamiltonian.

In order to verify the quality of this fit we define the error of energy for each converged case as

\vspace{-1em}
$$
\Delta E=|E_{\rm fit}-E_{\rm ab-initio}|
$$

Then, the maximum error $\mbox{max}(\Delta E)=1.1\ \mbox{meV}$, and the mean error $\langle\Delta E\rangle$ is equal to $0.44\ \mbox{meV}$.

Hamiltonian (\ref{eq:hiso}) is isotropic i.e. invariant under any rotation of spin vectors, however it implicitly 
contains magnetic anisotropy, because for a single angle $\alpha$ there appears a distribution of 
spherical angles $\theta_1$ and $\theta_2$, which correspond to different inclination angles $\beta_1$ and $\beta_2$ of magnetic moments of Cr atoms with respect to the hexagonal plane, $\beta=90^\circ-\theta$.

\begin{figure}[ht]
\begin{center}
\includegraphics[width=0.4\textwidth,trim={0 1.4cm 0 2.4cm},clip]{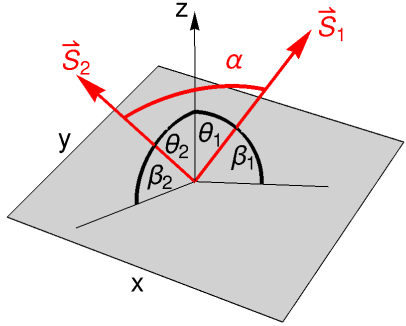}
\end{center}
\vspace{-1em}
\caption{Spherical angles $\theta_1$, $\theta_2$, inclination angles $\beta_1$, $\beta_2$,
 and mutual angle $\alpha$ for spin vectors $\vec{S}$ and $\vec{S}_2$.
\label{rysa}}
\end{figure}

\begin{figure}[ht]
\begin{center}
\includegraphics[width=0.4\textwidth]{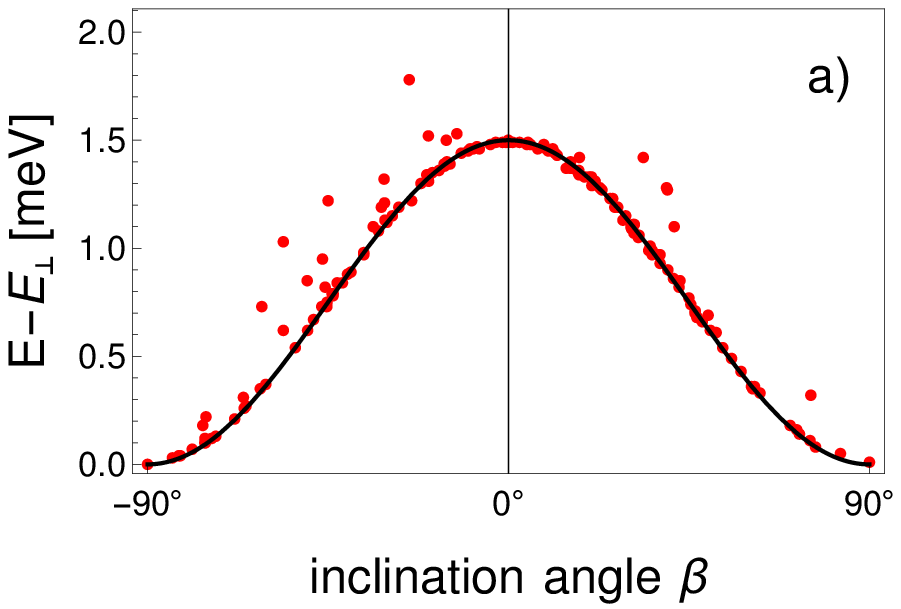}
\includegraphics[width=0.39\textwidth]{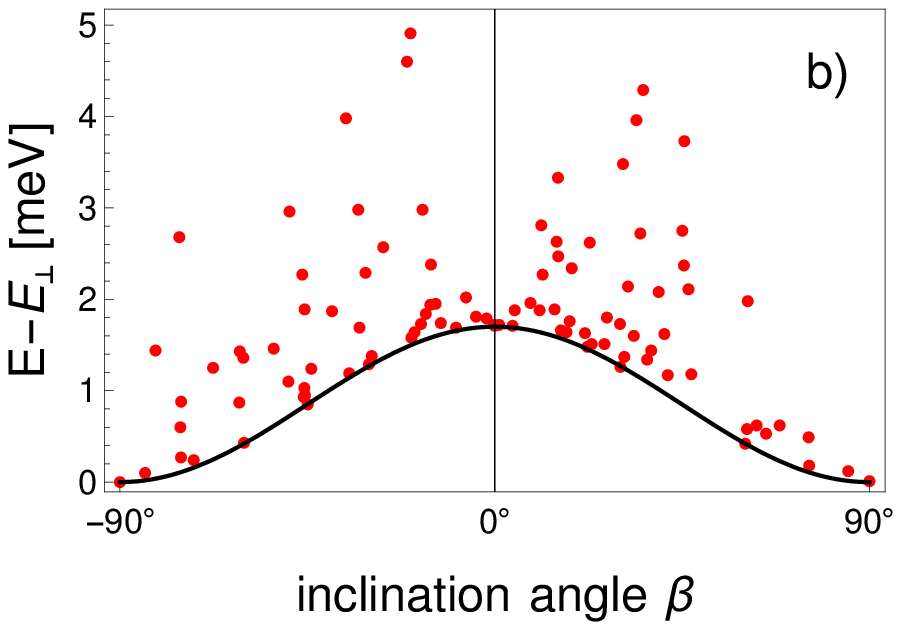}
\end{center}
\caption{Energy as a function of inclination angle $\beta$ for FM configurations: a) for $U=0\ \mbox{eV}$, b) for $U=3.5\ \mbox{eV}$.
\label{fig5}}
\end{figure}

Magnetic anisotropy i.e. the dependence of magnetic energy $E$ on inclination angle $\beta$ of magnetic moment with respect to the $(x,y)$ plane, can be explicitly determined for FM configurations with $\theta_1=\theta_2=\theta$,
by plotting $E$ as a function of $\beta$ (see Fig.~\ref{fig5}). For a simple quadratic model

\vspace{-1em}
\begin{equation}
E-E_\perp=A\cos^2\beta=A\big(1-\cos^2\theta\big)
\label{eq:EMAEA}
\end{equation}

by least-squares fit in the case of $U=0\ \mbox{eV}$ we obtain $A=1.5\ \mbox{meV}$, which results in $0.75\ \mbox{meV}$ per Cr atom in the unit cell.

For $U=3.5\ \mbox{eV}$ the functional dependence (\ref{eq:EMAEA}) is not clearly defined, because in this case
ab-initio magnetic structures correspond only approximately to FM order, as can be seen from the left panel in Fig.~\ref{fig3},
where there appear deviations from the straight line $\theta_1=\theta_2$.

\subsection{$\rm\textbf C_3$ and $SO(2)$-invariant Hamiltonians}

The isotropic quartic Hamiltonian (\ref{eq:hiso}) describes approximately the magnetic energy of $\rm CrI_3$
monolayer, however its errors of energy $\Delta E$ are of the order of weak magnetic anisotropy of this system.

To describe the dependence of total energy $E(\theta_1,\theta_2,\varphi_1,\varphi_2)$
of $\rm CrI_3$ monolayer on spherical angles of spin vectors $\vec{S}_1$ and $\vec{S}_2$ of Cr atoms
with a better phenomenological approximation,
we apply the following $C_3$-invariant Hamiltonian:

\begin{equation}
H=E_0+H_{\rm MCA}+3\big[H^{(2)}+H^{(2)}_{\rm quartic}\big]
\label{eq:Hquar}
\end{equation}

which is the sum of $H_{\rm MCA}$ quadratic (\ref{eq:mca2}) magnetocrystalline anisotropy with single parameter $A$ for two magnetic moments in the unit cell, quadratic anisotropic exchange $H^{(2)}$ given by Eq.~(\ref{eq:hs2}) with two parameters $J_1$ and $J_2$, and quartic term (\ref{eq:hc3})
with five parameters $K_1,\ldots\,K_5$. The last two terms are multiplied by 3 to take into account interaction of a given magnetic moment with three nearest neighbors. The non-magnetic constant energy $E_0$ is also treated as fitting parameter.
We do not take into account the quartic correction (\ref{eq:mca4}) to magnetocrystalline energy, because five parameters
of the quartic term $H^{(2)}_{\rm quartic}$ deliver enough free parameters to obtain a good fit.

We apply the linear least-squares method to fit all nine parameters at once to the results of 148 converged ab-initio calculations with $U=3.5\ \mbox{eV}$. We believe that this procedure is superior to that of the four-states energy mapping method \cite{Sabani2020} because: 1) we test the validity of Heisenberg model for all spherical angles of possible spin vectors. 2) We can directly verify the quality of assumed model by evaluating the difference between ab-initio and fitted energies for all converged cases.

In case of four-states energy mapping approach: 1) one only takes a few selected configurations, usually only in-plane or out-of plane, to determined value of each single exchange parameter,
assuming without proof that Heisenberg model works for any angle between spin vectors. 2) Because one solves $N=4$ equations 
with $N=4$ parameters the solution is exact, however there is no indication of uncertainty of obtained numerical values.

Least-squares method yields the values of $C_3$-invariant Hamiltonian parameters\footnote{We omit non-magnetic energy $E_0$.} for monolayer $\rm CrI_3$, presented in Tab.~\ref{tab1}, expressed in meV and rounded to two decimal digits.
The value of the anisotropy parameter $A$ is smaller than that determined from simple quadratic model (\ref{eq:EMAEA}) for $U=0\ \mbox{eV}$, due do the approximate nature of FM configurations for $U=3.5\ \mbox{eV}$. 
From Eq. (\ref{eq:mca2}) $E_{\rm MCA}=A\cos^2\theta$, hence opposite sign with respect to Eq.~(\ref{eq:EMAEA}).
Maximum error of energy $\mbox{max}(\Delta E)$ equals $0.58\ \mbox{meV}$, and the mean error is $0.12\ \mbox{meV}$.

\begin{table}
$$
\begin{array}{|r||r|r||r|r|r|r|r|}
\hline
A & J_1 & J_2 & K_1 & K_2 & K_3 & K_4 & K_5\\
\hline
-0.39 & -8.18 & -8.59 & -3.00 & 0.00 & 0.01 & -0.97 & -2.85 \\
\hline
\end{array}
$$
\caption{\label{tab1}Parameters of the $C_3$-invariant Hamiltonian for $\rm CrI_3$ monolayer expressed in meV.}
\end{table}

\begin{table}
$$
\begin{array}{|r||r|r||r|r|r|}
\hline
A & J_1 & J_2 & K_1 & K_2 & K_3 \\
\hline
-0.35 & -8.18 & -8.59 & -2.98 & -0.97 & -2.86 \\
\hline
\end{array}
$$
\caption{\label{tab2}Parameters of the $SO(2)$-invariant Hamiltonian for $\rm CrI_3$ monolayer expressed in meV.}
\end{table}

As seen in Tab.~\ref{tab1} parameters $K_2=K_3=0$, which confirms the assumption made in the previous section that one can apply more general $SO(2)$ symmetry for magnetic monolayers\footnote{Cf. comment after Eq. (\ref{eq:hso2}). In the second fit, parameter $K_1$ is multiplied by 3, in order to present the same values in both tables.}. We therefore re-evaluated linear least-squares fit replacing five-parameter 
quartic Hamiltonian (\ref{eq:hc3}) with $SO(2)$-invariant one (\ref{eq:hso2}), which has only three parameters.
The resulting fitted values, rounded to two decimal digits, are presented in Fig.~\ref{fig2}.
Maximum error of energy for $SO(2)$-invariant Hamiltonian $\mbox{max}(\Delta E)$ equals $0.65\ \mbox{meV}$, and the mean error is $0.12\ \mbox{meV}$. These errors are below the value of weak magnetocrystalline anisotropy for $\rm CrI_3$ monolayer, therefore
we believe that $SO(2)$-invariant Hamiltonian (\ref{eq:Hquar}) correctly describes the magnetic energy of this system.


\section{\label{secC}Conclusions}

Our results indicate that fourth-order corrections to anisotropic Heisenberg Hamiltonian are important to precisely describe the energy of magnetic configurations of two-dimensional magnetic materials. We believe that the $SO(2)$ symmetry generates the correct 
phenomenological description of magnetic structures suitable for application to diverse two-dimensional materials.
In conjunction with the idea of mean invariant tensor it allows to reduce the number of parameters of the model to the minimum.
The method of global least-squares fit to determine parameter values offers an advantage over four-state energy mapping, supplying the estimate of systematic energy errors, and allowing for verification of adequacy of Heisenberg model for arbitrary spin configurations.

The results of our work could be extended and applied to two-dimensional Janus magnetic materials \cite{Liang2020}, where the appearance of very large Dzyaloshinskii–Moriya interaction would possibly require going beyond standard description of DMI.


\begin{acknowledgments}
Numerical calculations were supported in part by PL-Grid Infrastructure. 
\end{acknowledgments}

\end{document}